\documentclass[aps,preprint]{revtex4}
\usepackage{graphicx}
\usepackage{bm}

\newcommand{\be}[1]{\begin{equation}\label{#1}}
\newcommand{\ee}{\end{equation}}
\newcommand{\ba}[1]{\begin{eqnarray}\label{#1}}
\newcommand{\ea}{\end{eqnarray}}

\def\la{\langle}
\def\ra{\rangle}

\begin{document}
\date{\today}
\title{Unified description of resonance and decay phenomena} 
\author{Ingrid Rotter}
\affiliation{Max Planck Institute for the Physics of Complex
Systems, D-01187 Dresden, Germany }

\begin{abstract}
In the Feshbach projection operator formalism, resonance as well as decay
phenomena are described by means of the complex
eigenvalues and eigenfunctions of the
non-Hermitian Hamilton operator  $H_{\rm eff}$
that appears in an intermediate stage of the formalism.
The formalism can be applied for the description of isolated resonances 
as well as for resonances in the overlapping regime.
Time asymmetry 
is related to the time operator which is a part of $H_{\rm eff}$.
An expression for the decay rates of resonance states 
is derived.  For isolated resonance states $\lambda$, this expression 
gives the fundamental relation 
$\tau_\lambda = \hbar / \Gamma_\lambda$
between life time and width of a resonance state. A similar relation holds for 
the average values obtained for narrow
resonances superposed by a smooth background term. 
In the cross over between these two cases
(regime of overlapping resonances), the decay rate decreases 
monotonously as a function of increasing time.  

\end{abstract}

\pacs{03.65.Ca, 03.65.Yz, 11.30.Er, 21.10.Tg}

\maketitle

\section{Introduction}

An old problem of standard quantum mechanics is the relation
between  the life time $\tau_\lambda$ of  
a resonance state and  its width $\Gamma_\lambda$ that determines
the Breit-Wigner energy distribution around the energy of 
this state in  a scattering process. 
The problem arises, on the one hand, from the fact that the states of 
standard quantum
mechanics are discrete, i.e. they are not coupled to the scattering states
of the continuum and have an infinite long life time. An exception are 
Gamow states being decaying states with a finite life time. 
They have however a single-particle structure and do not correspond to the
resonance states with more complicated many-particle structure
and longer life time. Moreover, they do not exist in Hilbert space quantum
mechanics. 
On the other hand, the spectroscopic information such as
position $E_\lambda$ in energy and width $\Gamma_\lambda$ of a resonance state
is obtained, usually, from the poles of the $S$ matrix, i.e. by continuing the
$S$ matrix into the complex energy plane.
The spectroscopic information obtained in this manner, can be considered to be 
reliable only for long-lived resonance states the widths of which are small.   
Thus, the relation between $\tau_\lambda$
and $\Gamma_\lambda$ of a resonance state can not be derived convincingly
in standard quantum mechanics. 

The present experimental situation is more convincing.
A few years ago, high precision measurements have been performed
\cite{volz,oates} in which life times and decay widths of resonance 
states have been measured
independently from one another. The experimental values confirm the relation
\begin{eqnarray}
\tau_\lambda = \hbar / \Gamma_\lambda
\label{tauga}
\end{eqnarray}
to a high degree of accuracy. Recently, a violation of the exponential decay
law at long times has been found experimentally in luminescence decays of many
dissolved organic materials after pulsed laser excitation over 
more than 20 life times \cite{rothe}. The turnover into the non-exponential
decay regime takes place sharply (in double-logarithmic scale) at long times.

As has been shown  by Bohm et al. \cite{bohm,bohmbook} (see also the 
recent papers \cite{bohm1,bohm2}), a consistent mathematical theory 
with  unification of resonance and decay phenomena
according to the relation (\ref{tauga})
leads to time asymmetry of quantum mechanics. 
While in the standard quantum theory the set of states and the set of
observables are mathematically identified and described by the same Hilbert
space, this is not so in the theory by Bohm et al. Instead, the Hardy space  
is introduced with time asymmetrical boundary conditions
for time symmetric dynamical equations. 
According to this theory, time
asymmetry is inherent in the dynamics of a quantum system. One distinguishes
between prepared states and detected observables: the observables can be
detected only {\it after} the states are prepared. 
In the framework of this theory, a simple 
description of decaying states can be given for Gamow states.
The description of resonance states with more
complicated structure remains however a problem.

Another possibility to unify resonance and decay phenomena is to consider
a non-Hermitian Hamilton operator instead of the Hermitian Hamilton operator
basic of standard quantum mechanics. The eigenvalues of a non-Hermitian
operator are complex and provide therefore the life times of resonance states
in a natural manner via the relation (\ref{tauga}). 
Numerical studies have been performed by using two different methods. One of
these
 methods is based on the Feshbach projection operator (FPO) technique
\cite{ro91rep} while the other one uses complex scaling \cite{moisrep}.

Using the FPO technique, 
the decay rates of resonance states are calculated 
\cite{decayrate} some years ago for isolated as well as for overlapping
resonance states. In these calculations,
the resonance states are described by the eigenstates of a 
non-Hermitian Hamilton
operator. While the decay rates are constant in time as long as the resonance
states do not overlap, they start to oscillate in the neighborhood of
branch points in the complex plane where the eigenvalues of two resonance
states coincide. These oscillations are caused by the fact that the decay rate 
at the considered energy $E$ is determined not only by the life time of one 
individual resonance state but also by that of the neighboring one
with which it overlaps. 
That means, the decay rate of an individual resonance state is
ill defined in the regime of overlapping resonance states.
Accordingly to this result, it holds $\Xi_\lambda(t) = \Phi_\lambda^*(t)$
for the dual basis vectors $\Xi_\lambda$ and $\Phi_\lambda$ of  a
complex symmetric non-Hermitian Hamilton operator in the neighborhood of the
branch point (exceptional point)  only at the time $t=t_0$ \cite{gurosa}.
Taking into account the contributions of all overlapping resonance states
at the energy $E$, the decay rate ceases to oscillate and decreases 
smoothly as a function of time \cite{decayrate}.

Complex scaling is used in
\cite{moiseyev} in order to calculate time dependent observables.
In this paper, the time-asymmetry
problem in non-Hermitian quantum mechanics is discussed. It is argued that
the non-Hermitian description of the system is valid on time scales that are
long enough to regard only the localized part of the wave packet after the
scattered part has left the interaction region. In the numerical calculations,
isolated resonances are
considered, for which the decay rate does not oscillate.

In other approaches, the quantum measurement and its 
relation to the arrow of time is considered, see e.g. Schulman \cite{schulman}.
Recently, the passage time distribution for a spread-out quantum particle to
transverse a specific region is calculated using a special quantum model for
the detector involved \cite{hegerfeldt}. In this model, the coupling between a
collection of spins and their environment is enhanced by the detection
of the particle. Such a model is not in contradiction with
the results of the FPO method, according to which  an
enhancement of the coupling between system and environment will, generally, 
be generated in the regime of overlapping resonance states due to the
alignment of the resonance states to the states of the environment
\cite{brscorr,robra}. In the FPO formalism, the enhancement of the coupling 
is, however, independent
of whether or not the particle is detected. It is an internal property of an
open quantum system

In the present paper, it will be shown that preparation
and detection of resonance states are both involved also in the theory 
based on the FPO method. Thus, this method embodies  the time
asymmetry of quantum mechanics studied by Bohm et al. \cite{bohm,bohmbook}.
It is involved in the non-Hermitian part of the
Hamilton operator $H_{\rm eff}$ that describes 
the resonance states in the FPO formalism and contains the time operator. 
Moreover, the FPO formalism allows  also to study  time asymmetry in a 
system with narrow  resonance states (different from the  Gamow 
states) and of resonance states near decay thresholds and
in the overlapping regime. 

In Sect. II of the present paper,
the basic equations of the FPO formalism are given, including the
expression for the resonance part of the $S$ matrix. The peculiarities of
the FPO formalism are summarized and contrasted with the  standard 
formalism in Sect. III. In the following Sect. IV, an expression for the 
time dependent decay rate is given that holds true for isolated as well as
for overlapping resonance states. In Sect. V, the decay rate in the
regime of overlapping resonances is considered in detail. 
The results are summarized in the last section.

\section{Basic relations of the Feshbach projection operator formalism }

In the FPO formalism \cite{feshbach}, 
the full function space is divided into two subspaces:
the $Q$ subspace contains all wave functions that are localized inside the 
system and vanish outside of it while
the wave functions of the $P$ subspace are extended up to infinity and vanish
inside the system, see \cite{ro91rep}. 
The wave functions of the two subspaces can be obtained by standard methods: 
the $Q$ subspace is described by the Hermitian Hamilton operator $H_B$
that characterizes the  closed system with discrete states, while
the $P$ subspace  is described by the Hermitian Hamilton
operator $H_C$ that contains the continuum of scattering wave functions.
In the FPO formalism, the closed system (defined by the 
Hamilton operator $H_B$) will be opened by coupling the wave functions of the 
$Q$ subspace to those of the $P$ subspace under the assumption $P+Q=1$.
Due to this coupling, the discrete states of the closed system pass into
resonance states of the open system. The resonance states have,
in general, a finite life time. 

The basic equation of the FPO formalism  
\begin{eqnarray}
(H-E)\;\Psi^E_C = 0 
\label{Psi}
\end{eqnarray}
has to be solved in the whole function space $P+Q$.
It contains the decay of the subsystem  localized in the $Q$ subspace, 
into the surrounding $P$ subspace where the decay products will be detected. 
The excitation of the states localized in the $Q$ subspace may take place 
via one of the channels $C$ included in (\ref{Psi}) or by another
process described by
\begin{eqnarray}
(H-E)\;\Psi^E_F = F,
\label{PsiF}
\end{eqnarray}
where the inhomogeneity (source term) $F$ on the 
right-hand side of (\ref{PsiF})
describes the excitation  of the state $\Psi^E_C$ by a process
different from scattering. It may describe, e.g., the Coulomb excitation of 
nuclear states in photo-nuclear reactions, see \cite{Barz}. 
In this case, $F=H_{\rm int} \phi_T$ where $H_{\rm int}$ is the interaction
of the electromagnetic field with the target ground state $\phi_T$. 
Eq. (\ref{PsiF})  may describe also the excitation of
an optically prepared sample of ultra-cold atoms. In the present paper, 
the value $F$ will not be specified. It is $F=0$ only in the scattering
process. It should be underlined however that
both equations (\ref{Psi}) and (\ref{PsiF}) are defined in the whole
function space defined by $P+Q=1$. 
Therefore, the Hamilton operator $H$ appearing in these
equations  is  Hermitian.

 In solving (\ref{Psi}) and (\ref{PsiF}) in the  function space 
$P+Q=1$ by using the 
FPO technique, the  non-hermitian Hamilton operator 
\begin{eqnarray}
H_{\rm eff} = H_B + \sum_C V_{BC} \frac{1}{E^+ - H_C} V_{CB} 
\label{heff}
\end{eqnarray}  
appears which contains $H_B$ as well as an additional non-hermitian term 
that describes the coupling of the resonance states via the common environment.
Here  $V_{BC}, ~V_{CB}$ stand for the coupling matrix elements between the 
{\it eigenstates} of $H_B$ and the environment \cite{ro91rep}
that may consist of different continua $C$. The operator $H_{\rm eff}$
is symmetric,
\begin{eqnarray}
(H_{\rm eff} - z_\lambda)\,\phi_\lambda =0 \; ,
\label{phi}
\end{eqnarray}
its eigenvalues $z_\lambda$ and eigenfunctions 
$\phi_\lambda$ are complex. The eigenvalues
provide not only the energies of the resonance
states but also their widths. The eigenfunctions
are biorthogonal. For details see \cite{ro91rep}.

The eigenvalues and eigenfunctions of $H_B$ contain the  
interaction $u$ of the discrete states which is given by the 
nondiagonal matrix elements of $H_B$. This interaction 
is of standard type in closed systems and may be called therefore
internal interaction. The
eigenvalues and eigenfunctions of $H_{\rm eff}$ contain additionally the
interaction $v$ of the resonance states via the 
common continuum ($v$ is used here instead of the concrete 
matrix elements of the second term of $H_{\rm eff}$). 
This part of interaction is, formally, of second order and
may be called  external interaction.
While $u$ and Re$(v)$ cause  level repulsion in energy, 
Im$(v)$ is responsible for the bifurcation of the widths 
of the resonance states  (resonance trapping). 
The phenomenon of resonance trapping
has been proven experimentally in microwave cavities \cite{stm}.

Since the effective Hamilton operator (\ref{heff}) depends explicitly  on 
energy $E$, so do its eigenvalues $z_\lambda$ and eigenfunctions
$\phi_\lambda$. Far from thresholds, the energy dependence 
is weak, as a rule, in an energy interval of the order of magnitude of the 
width of the resonance state. 
The solutions of the fixed-point equations
$E_\lambda={\rm Re}(z_\lambda)_{|E=E_\lambda}  $ and of
$\Gamma_\lambda=-2\, {\rm Im}(z_\lambda)_{|E=E_\lambda} $
are numbers that coincide  with the poles of the $S$ matrix. 
In the FPO formalism, however, it is not necessary to consider the
poles of the $S$ matrix since the spectroscopic information 
on the system follows directly from the complex
eigenvalues $z_\lambda$  of $H_{\rm eff}$. Moreover,
in the physical observables related to the  $S$ matrix  
the eigenvalues $z_\lambda$ with their full energy dependence are involved,
see (\ref{smatr}).  Due to this fact,
information on  the vicinity
(in energy) of the considered resonance states such as the position of decay
thresholds and of neighboring resonance states is involved in the $S$ matrix
and can be received. Such an information can not be
obtained from the poles of the $S$ matrix being (energy-independent) numbers. 

In contrast to the (parametric) trajectories of the 
eigenvalues of a Hermitian Hamilton operator, those of a
non-Hermitian one may cross.
The crossing points are branch points in the complex energy plane (called
exceptional points in the mathematical literature). Physically, they are
responsible for the avoided level crossing phenomenon appearing in their
vicinity. More precisely:
in approaching the branch points under different conditions, we have 
level repulsion (together with widths equilibration) or widths bifurcation
(together with level attraction). For details see \cite{ro91rep}.  

The eigenfunctions $\phi_\lambda$  of $H_{\rm eff}$
are complex and  biorthogonal,
\begin{eqnarray} 
\langle\phi_\lambda^*|\phi_{\lambda '}\rangle = \delta_{\lambda, \lambda '} 
\label{biorth1} 
\end{eqnarray}
with the consequence that \cite{ro91rep}
\begin{eqnarray}
\langle\phi_\lambda|\phi_{\lambda}\rangle & \equiv & A_\lambda \ge 1
\label{biorth2a} \\
B_\lambda^{\lambda '} \equiv 
\langle\phi_\lambda|\phi_{\lambda ' \ne \lambda}\rangle & = & -B_{\lambda
'}^\lambda \equiv  
- \,\langle\phi_{\lambda '\ne \lambda }|\phi_{\lambda }\rangle 
\nonumber \\
&&
|B_\lambda^{\lambda '}| ~\ge ~0  \; .
\label{biorth2b}
\end{eqnarray}
The normalization condition (\ref{biorth1}) entails that 
the phases of the eigenfunctions in the overlapping regime
are not rigid: the normalization condition
$\la\phi_\lambda^*|\phi_{\lambda}\ra =1$ is fulfilled
only when Im$\langle \phi_\lambda^*|\phi_\lambda\rangle \propto $
Re$~\phi_\lambda \cdot$  Im$~\phi_\lambda =0$, i.e. in the regime of
overlapping resonances
by rotating the wave function at a certain angle $\beta_\lambda$. 
For details see \cite{robra}.

The solution of (\ref{Psi}) reads \cite{ro91rep}
\begin{equation}
\label{total}
|\Psi_C^E \rangle= |\xi^E_C\rangle + \sum_{\lambda}
|\Omega_\lambda^C \rangle
~\frac{\langle\phi_\lambda^* |V| \xi^E_C\rangle}{E-z_\lambda} 
\end{equation}
where
\begin{equation}
\label{reswf}
|\Omega_\lambda^C \rangle= 
\Big( 1+ \frac{1}{E^{+}-H_C} V_{CB}\Big)|\phi_\lambda \rangle 
\end{equation}
is the wave function of the resonance state $\lambda$ and
the $\xi^E_C$ are the (coupled) scattering wave functions 
of the continuum into which the system is embedded.
According to (\ref{total}), the eigenfunctions $\phi_\lambda$ of the 
non-Hermitian Hamilton operator $H_{\rm eff}$ give the main
contribution to the scattering wave function $\Psi^E_C$ in the interior 
of the system,
\begin{eqnarray}
|\Psi_C^E \rangle \to |\hat \Psi_C^E \rangle =
\sum_\lambda c_{C \lambda}^E\, |\phi_\lambda \rangle  \, ; 
\quad \; c_{C \lambda}^E =
\frac{\langle \phi_\lambda^* |V| \xi^E_C\rangle}{E-z_\lambda } 
\label{total1}
\end{eqnarray}
and 
\begin{eqnarray}
\langle\Psi_C^E | \to \langle\hat \Psi_C^E | =
\sum_\lambda c_{C \lambda}^{E*} \, \langle\phi_\lambda^{\rm left} |  
= \sum_\lambda c_{C \lambda}^{E*} \, \langle\phi_\lambda^{*} | \; .
\label{total1l}
\end{eqnarray}
The weight factors $c_{C \lambda}^{ E}$
contain the decay  of the states $\lambda$ at the energy $E$.
The solution of (\ref{PsiF}) is \cite{Barz,ro91rep}
\begin{eqnarray}
|\Psi_F^E \rangle &=& 
|\xi^E_F\rangle + \sum_{\lambda}
|\Omega_\lambda^C \rangle 
~\frac{\langle \Omega_\lambda^{C*}|F\rangle}{E-z_\lambda} 
\nonumber \\ &=& 
|\xi^E_F\rangle + \sum_{\lambda}
|\Omega_\lambda^C \rangle
~\frac{\langle\phi_\lambda^* |(V_{BC}~[E-H_C]^{-1}~P +Q)F\rangle}{E-z_\lambda} 
\label{totalF}
\end{eqnarray}
where $(E-H_C)\xi_F^E=PF$. In the interior of the system, it is
\begin{eqnarray}
|\Psi^E_F\rangle \to |\hat\Psi^E_F\rangle 
= \sum_\lambda c_{F \lambda}^E |\phi_\lambda \rangle \, ;
\quad \, c_{F \lambda}^E =
\frac{\langle \phi_\lambda^* | QF \rangle}{E-z_\lambda}
\label{total1F}
\end{eqnarray}
and
\begin{eqnarray}
\langle\Psi^E_F | \to \langle\hat\Psi^E_F | 
= \sum_\lambda c_{F \lambda}^{E*} \langle\phi_\lambda^{\rm left} | 
= \sum_\lambda c_{F \lambda}^{E*} \langle\phi_\lambda^{*} | 
\; .
\label{total1Fl}
\end{eqnarray}

The $S$ matrix follows from $\langle \xi^E|V|\Psi^E_C\rangle $  
~(where $\xi^E$ stands for  $\xi^E_C ~{\rm or} ~\xi^E_F$), see \cite{ro91rep}. 
The amplitude of the resonance part we are interested in, is given by
\begin{eqnarray}
S^{\rm res} = i \sum_\lambda \langle\xi^E|V|\phi_\lambda\rangle 
~c^E_{C \lambda}
=i\sum_\lambda\frac{ \langle\xi^E|V|\phi_\lambda\rangle \langle
  \phi_\lambda^*|V|\xi^E_C\rangle }{E-z_\lambda} \; .
\label{smatr}
\end{eqnarray}
This expression shows immediately that the resonance phenomena (described 
by the $S$ matrix) are determined by the decay properties of the resonance 
states (described by the complex eigenvalues $z_\lambda$ 
and eigenfunctions $\phi_\lambda$ of the
non-Hermitian Hamilton operator $H_{\rm eff}$). Thus, the FPO formalism 
provides a unified description of resonance and decay phenomena. 
The expression (\ref{smatr}) shows however also that, generally, 
the energy dependence of the
eigenvalues $z_\lambda$ and eigenfunctions $\phi_\lambda$  of $H_{\rm eff}$
causes deviations from the Breit-Wigner resonance line shape
and from the exponential decay law. The deviations
become important for isolated resonance
states in the long-time scale due to the fact that the decay
thresholds lie at a finite energy \cite{cusp}. 
This result agrees qualitatively with experimental data \cite{rothe}.
At high level density, deviations appear even in the short-time 
scale  due to the mutual influence of neighbored resonance states, 
see  Sect. V and \cite{robra}.

\section{Peculiarities of the Feshbach projection operator formalism }

The main advantages of the FPO formalism consist in the following.

(i) The spectroscopic information on the resonance states is
obtained directly from the complex eigenvalues $z_\lambda$
and eigenfunctions $\phi_\lambda$
of the non-Hermitian Hamilton operator $H_{\rm eff}$.
The $z_\lambda$ and $\phi_\lambda$
are energy dependent functions, generally, and contain the
influence of neighboring resonance states as well as of decay thresholds
onto the considered state $\lambda$. This energy dependence allows to describe 
decay and resonance phenomena also in the very neighborhood of
decay thresholds and in the regime of overlapping resonances.
Since also the coupling coefficients between system and continuum 
depend on energy,  the unitarity of the $S$ matrix is guaranteed, 
see e.g. \cite{ro03}.

(ii) The resonance states are directly related to the discrete states 
of a closed system described by standard quantum mechanics (with the Hermitian
Hamilton operator $H_B$). They  are generated by opening the system, i.e. by
coupling the discrete states to the environment of scattering states by 
means of the second term of
the Hamilton operator $H_{\rm eff}$. Therefore, they are realistic
(long-lived many-particle)  states of an  open quantum system.

(iii) In the FPO formalism it
is not necessary to consider the poles of the $S$ matrix. Therefore,
additional mathematical problems in the neighborhood of branch points 
(exceptional points) in the complex plane are avoided.

(iv) The phases of the eigenfunctions $\phi_\lambda$ 
of $H_{\rm eff}$ are not rigid in the vicinity of a branch point.
This fact allows spectroscopic reordering processes in the system  
under the influence of the scattering wave functions of the environment into
which the system is embedded.

These features are involved in all present-day \cite{presentday} 
calculations performed on the 
basis of the FPO formalism. In numerical studies, 
the main problem arises from the definition of the
two subspaces $Q$ and $P$ such that it is meaningful for spectroscopic
studies (see the discussion of this point in the reviews \cite{ro91rep}).
The basic idea is the following: $H_B$ describes the closed system
(localized in the interior of the system)
which becomes open when embedded into the environment of 
the extended scattering wave
functions described by $H_C$. Therefore, all values characteristic of
resonance states can be traced back
to the corresponding values of discrete states
by controlling the coupling to the continuum. That means with $v\to 0$, 
the transition from resonance states (described by the non-Hermitian 
$H_{\rm eff}$) to discrete states (described by the Hermitian $H_B$)
can be controlled.  

Another peculiarity of the FPO formalism is the existence of
a time operator which is the residuum  of the non-Hermitian Hamilton
operator $H_{\rm eff}$. The life time $\tau_\lambda$ of a resonance state
follows from the eigenvalue $z_\lambda$ of 
$H_{\rm eff}$ in the same manner as the energy $E_\lambda$ of
this state. Both values are fundamentally different from the time $t$ and the
energy $E$. They characterize the states $\lambda$ while $t$ and
$E$ appear as general parameters. In the closed system with the 
Hermitian Hamilton operator $H_B$, only the energies $E_B$ of the states can be
determined. The eigenvalues are real and the widths are zero, $\Gamma_B =0$.
Due to the coupling to the continuum, energy shifts 
$E_\lambda - E_B$ of the states
appear as well as the finite life times $\tau_\lambda
\propto (\Gamma_\lambda - \Gamma_B)^{-1} = 
\Gamma_\lambda^{-1} $ of the resonance states. 
Both, the energy shifts and the finite life times, 
follow from the second term of the non-Hermitian operator
$H_{\rm eff}$ [see Eq. (\ref{heff})].
Usually, the numbers $E_\lambda $  and $\Gamma_\lambda$
can  be obtained directly from the $z_\lambda$.
Only in the case the $z_\lambda $ are strongly dependent on energy,
the corresponding fixed-point equations have to be solved.
The energies $E_\lambda$ and  life times $\tau_\lambda$ of the resonance 
states $\lambda$ of an open quantum system are bounded from below
(see \cite{robra} for the discussion of the
brachistochrone problem in open quantum systems). 
Mathematically, the existence of the time operator entails
the time asymmetry involved in the FPO formalism.

\section{Time dependent values}

The time dependent Schr\"odinger equation reads
\begin{eqnarray}
H_{\rm eff} ~\hat\Psi^E(t)=i~\hbar ~\frac{\partial}{\partial t}
~\hat\Psi^E(t) \, .
\label{tdse0}
\end{eqnarray}
The right solutions may be represented, according to (\ref{total1}), 
by an ensemble of resonance states $\lambda$ that describes the decay
of the system  at the energy $E$, 
\begin{eqnarray} 
|\hat\Psi^{E~\rm (right)}(t)\rangle & =& 
e^{-iH_{\rm eff} \, t/\hbar} ~|\hat\Psi^{E~\rm (right)}(t_0)\rangle 
\nonumber \\
&=& \sum_{\lambda } ~e^{-i z_{\lambda } \, t/\hbar}  
~c_{\lambda 0}
~|\phi_{\lambda }^{\rm (right)}\rangle 
\label{tdse1}
\end{eqnarray}
with $|\phi_{\lambda }^{\rm (right)}\rangle =|\phi_{\lambda }\rangle$
and  $c_{\lambda 0} = 
\langle\phi_\lambda^* |V|\xi^E_C\rangle/(E-z_\lambda) $.
The $z_\lambda$ and $\phi_\lambda$ are the (energy dependent) 
eigenvalues and eigenfunctions of
the time-independent Hamilton operator $H_{\rm eff}$, Eq. (\ref{heff}),
while the $\xi^E_C$ are the scattering wave functions of the environment.
The left  solution  of (\ref{tdse0}) reads
\begin{eqnarray} 
\langle \hat\Psi^{E~\rm (left)}(t)| & = & 
\langle\hat\Psi^{E~\rm (left)}(t_0)|  ~e^{iH_{\rm eff}^{\dagger} \, t/\hbar} 
\nonumber  \\
& = & \sum_{\lambda } 
~\langle\phi_{\lambda }^{\rm (left)}|  
~d_{\lambda t} ~e^{i z_{\lambda }^* \, t/\hbar} 
\label{tdse2}
\end{eqnarray}
with $\langle\phi_{\lambda }^{\rm (left)}| =\langle\phi_{\lambda }^{*}|$  and
$d_{\lambda t}  
=c_{\lambda 0}^{*}
=\langle \xi^E_C|V|\phi_\lambda\rangle/(E-z_\lambda^*)$ 
or $d_{\lambda t} = c_{F\, \lambda}^{E*}(t) = 
\langle F^*(t) Q|\phi_\lambda\rangle\, / \, (E-z_\lambda^*) $
according to (\ref{total1l}) and (\ref{total1Fl}), respectively.
It describes the excitation of the system at the energy $E$. The source term
is, generally, time dependent: $F=F(t)$.

By means of (\ref{tdse1}) and (\ref{tdse2}) the population probability 
\begin{eqnarray} 
\langle \hat\Psi^{E~\rm (left)}(t)|\hat\Psi^{E~\rm (right)}(t)\rangle = 
\sum_\lambda
c_{\lambda 0} ~d_{\lambda t} ~e^{- \Gamma_{\lambda } t/\hbar} 
\label{tdse3}
\end{eqnarray}
at the energy $E$ can be defined.
The decay rate reads 
\begin{eqnarray} 
k_{\rm gr}(t)& =& - \frac{\partial}{\partial t} ~{\rm ln} 
\, \langle \hat\Psi^{E~\rm (left)}(t) | \hat\Psi^{E~\rm (right)}(t) \rangle 
\nonumber \\
& =& \frac{1}{\hbar}~\frac{\sum_\lambda \Gamma_\lambda 
~c_{\lambda 0} ~d_{\lambda t}
~e^{- \Gamma_{\lambda } t/\hbar}}{\sum_\lambda 
c_{\lambda 0} ~d_{\lambda t} ~e^{- \Gamma_{\lambda } t/\hbar}} \; .
\label{tdse4}
\end{eqnarray}
For an isolated  resonance state $\lambda$,  (\ref{tdse4}) 
passes into the standard expression
\begin{eqnarray}
k_{\rm gr}(t) ~\to ~k_\lambda ~= ~\Gamma_{\lambda }/\hbar 
\; .
\label{kiso}
\end{eqnarray}
The value $k_\lambda$ is constant in time and  corresponds to 
(\ref{tauga}) with $\tau_\lambda = 1/k_\lambda $.
It describes the idealized case of an exponential decay law and,
according to (\ref{smatr}),
a Breit-Wigner resonance in the cross section.
Generally, deviations from the exponential decay law and from 
the Breit-Wigner line shape appear under the influence of neighboring
resonance states and (or) of decay thresholds  (see e.g. \cite{ro91rep}). 
Also the background term appearing in most reactions may cause deviations
from the ideal exponential decay law. 

The excitation process may occur on a much shorter time scale than the 
decay process. In such a case, the
function $d_{\lambda t}  $ will be a step-like function
at $t = t_0$. It is possible therefore to
study the pure decay process starting at  the time $t_0$.  When, 
in other cases,  $\langle \phi^*_\lambda|QF(t)\rangle $  is constant 
for $t > t_0$ (or increases with $t$ in a certain time interval), 
excitation  and decay take place at the same time in this time interval.    
This is the case also for
the scattering process corresponding to $F=0$.

Eq. (\ref{tdse4}) describes the decay rate also in the regime of overlapping
resonances. For numerical results see Ref. \cite{decayrate}.
The overlapping and
mutual influence of resonance states is maximal at the branch points in
the complex plane where two  eigenvalues $z_{\lambda}$ and
$z_{\lambda '}$ of the effective Hamilton operator $H_{\rm eff}$ coalesce.
Nevertheless, the decay rate is everywhere smooth as
can be seen also directly from (\ref{tdse4}).
This result coincides with the general statement 
according to which all observable
quantities behave smoothly at  singular points. 

The expressions (\ref{tdse1}) and (\ref{tdse2}) are valid only  when   
(\ref{total1}) holds, i.e. at  times $t$ at which the 
wave functions $\Psi^E $  have a localized part  
in the interior of the system at the energy $E$ so that the representation
(\ref{total1}) is meaningful at this energy. According to (\ref{tdse1})
and (\ref{tdse2}), this is the case for times
$t\ge t_0$ where $t_0$  is  a finite
value. Without loss of generality, it can be chosen $t_0 =0$. 
The quantum system described in the framework of the FPO formalism is
therefore time asymmetric. The time asymmetry is involved in 
the non-Hermitian part of the Hamilton operator $H_{\rm eff}$
which contains the time operator. This can be seen also from the
expression (\ref{tdse3}) for the population probability.

The consideration of only the time interval $0\le t \le \infty$ in 
(\ref{tdse0}) is related to the fact that the decay of a resonance state
(at the energy $E$ of the system) starts at a finite time 
(say $t_0=0$) at which the system
can be considered to be excited, i.e. (\ref{total1}) is meaningful
at this energy. This fact agrees with the concept of a semigroup description 
introduced in \cite{bohm},  
which distinguishes between prepared and measured states.  
In our formalism, the decaying (measured) states are described by the  
eigenvalues and eigenfunctions of the  non-Hermitian 
Hamilton operator $H_{\rm eff}$ involved in the $|\hat\Psi^E \rangle $,
Eq. (\ref{total1}). 
The preparation of the resonance states is described by the 
$\langle \hat \Psi^E |$, Eq. (\ref{total1Fl}).
It may be very different for different reactions. 

The decay properties of the resonance states
can be studied best when their excitation 
takes place in a time interval that is very short 
as compared to the life time $\tau_\lambda$ of the resonance states. In such a
case, the time $t_0=0$ is well defined and no perturbation of the decay process
by the still continuing excitation process will take place. 
In \cite{bohm2}, such a situation is studied in single ion experiments.
The results demonstrate  the beginning of time for a decaying state. 
That means, they prove the time asymmetry in quantum physics.

\section{Decay rates in the regime of overlapping  resonance states}

In the regime of overlapping resonance states,
spectroscopic reordering processes take place \cite{ro91rep}. 
Most interesting is the phenomenon of
width bifurcation and the loss of the phase rigidity of the wave functions
of the resonance states under the influence of the branch points in the
complex energy plane. For a detailed study of
the last phenomenon (loss of phase rigidity) see \cite{robra}.

The decay rate $k_{\rm gr}$, Eq.  (\ref{tdse4}),  
contains the widths $\Gamma_\lambda $ of the individual resonance states 
$\lambda$. When the resonance states overlap, it is however difficult 
to receive   information on the   
decay rates $k_\lambda(t)$ of the individual states. 
The reason is that (\ref{tdse4}) 
contains also the contributions from all the neighboring states 
$\lambda ' \ne \lambda $  at the energy $E$.
In order to get $k_\lambda(t)$, one has to consider 
\begin{eqnarray}
H_{\rm eff} ~\phi_\lambda(t)=i~\hbar 
~\frac{\partial}{\partial t}~\phi_\lambda(t) 
\label{tdse0a}
\end{eqnarray}
instead of (\ref{tdse0}) with 
\begin{eqnarray} 
|\phi_\lambda(t)\rangle & =& 
e^{-iH_{\rm eff} \, t/\hbar} ~|\phi_\lambda(t_0)\rangle \nonumber \\
&= &
~e^{-i z_{\lambda } \, t/\hbar}  
~c_{\lambda 0} ~|\phi_{\lambda }\rangle +
\sum_{\lambda ' \ne \lambda} 
~e^{-i z_{\lambda '} \, t/\hbar}  
~c_{\lambda ' 0} ~|\phi_{\lambda '}\rangle 
\label{tdse1a}
\end{eqnarray}
and
\begin{eqnarray} 
\langle \phi_\lambda(t)|& = &  ~\langle\phi_\lambda(0)|  
~e^{iH_{\rm eff}^\dagger \, t/\hbar} \nonumber \\
&=& 
~\langle\phi_{\lambda }|  
~~d_{\lambda t}
~e^{i z_{\lambda }^* \, t/\hbar} +
\sum_{\lambda '\ne \lambda}
~\langle\phi_{\lambda '}|  
~~d_{\lambda ' t}
~e^{i z_{\lambda '}^* \, t/\hbar} \; . 
\label{tdse2a}
\end{eqnarray}
The functions $\phi_\lambda$ are biorthogonal, 
see (\ref{biorth1}), (\ref{biorth2a}) and (\ref{biorth2b}). It follows
\begin{eqnarray} 
\langle \, \phi_\lambda(t)\, |\, \phi_\lambda(t)\,\rangle & = & 
c_{\lambda 0} \, d_{\lambda t} ~e^{- \Gamma_{\lambda }\, t/\hbar}\; 
\langle \phi_\lambda|\phi_\lambda \rangle
+  \nonumber \\
&& 
\hspace*{-2.5cm}
\sum_{\lambda \ne \lambda '}
e^{-(\Gamma_\lambda + \Gamma_{\lambda '})\, t/(2\hbar)} 
~\Big(
~c_{\lambda 0} d_{\lambda ' t} ~e^{i(E_{\lambda '} - E_{\lambda })\, t/\hbar}
- ~c_{\lambda ' 0} d_{\lambda t}~e^{i(E_{\lambda } - E_{\lambda '})\, t/\hbar}
\Big)
~\langle\phi_{\lambda '} | \phi_{\lambda }\rangle 
\label{tdse3a}
\end{eqnarray}
and  $k_{\rm \lambda}(t) = - \frac{\partial}{\partial t} ~{\rm ln} 
\langle \phi_\lambda(t) \, |\, \phi_\lambda(t) \rangle $
contains oscillating terms in the overlapping regime
at the energy $E$ considered. These oscillating terms 
correspond to the fact that, at a certain energy $E$ of the system,
an individual level is ill defined because of its overlapping with other
levels. The oscillations vanish 
by summing over the contributions from all individual states 
and considering the wave functions $\Psi^E$ that are solutions of 
(\ref{Psi}) with the Hermitian Hamilton operator $H$, see   
(\ref{tdse3}).

Thus, the oscillations of 
the decay rates $k_\lambda(t)$ of the individual resonance states 
illustrate in a direct manner how
neighboring resonance states influence one another. 
The physically relevant expressions for the decay rate are, however, the 
$k_{\rm gr} (t)$. For numerical results obtained for both values,
$k_{\rm gr} (t)$ and $k_{\lambda} (t)$,
in some special cases in the neighborhood
of branch points in the complex plane see \cite{decayrate}. 

Another interesting problem is the saturation of the average decay rate 
$k_{\rm av}$ in the regime of 
strongly overlapping resonances. According to the bottle-neck picture of the
transition state theory, it starts at a certain critical value of
bound-continuum coupling \cite{miller}. This saturation is caused by
widths bifurcation  (resonance trapping \cite{ro91rep}) occurring
in the neighborhood of the branch points in
the complex plane \cite{comment}. Widths bifurcation creates long-lived 
resonance states together with a few short-lived resonance states.
The definition of an average life time 
of the resonance states is meaningful therefore only 
for either the long-lived states or the short-lived ones. 
The long-lived (trapped) resonance states are almost decoupled from the
continuum of decay channels. Their  widths $\Gamma_\lambda$ 
saturate therefore with increasing bound-continuum coupling.
The $\Gamma_\lambda$ are almost the same for all  the different states 
$\lambda$, see \cite{ro91rep}, i.e.
$\Gamma_{\rm av} \approx \Gamma_\lambda$ for all long-lived trapped
resonance states. It follows therefore 
\begin{eqnarray}
k_{\rm av} 
\approx \Gamma_{\rm av} /\hbar 
\label{kapav}
\end{eqnarray}
from (\ref{tdse4}). According to the average width $\Gamma_{\rm av}$,
the average life time  of the long-lived states can be defined by 
$\tau_{\rm av} = 1/k_{\rm av} $. Then (\ref{kapav}) is equivalent 
to the basic equation (\ref{tauga}). That means, the basic relation
between life times and decay widths of resonance states holds 
not only for isolated resonance states [see Eq. (\ref{kiso})], but
also for narrow resonance states superposed by a smooth background
(originating from a few short-lived resonance states \cite{ro91rep,brscorr}).
In the last case, the relation holds for the average values 
$\Gamma_{\rm av}$ and $\tau_{\rm av}$.

\section{Summary}

As has been discussed in this paper, the FPO formalism is characterized by two
Hamilton operators: the Hermitian $H$ and the non-Hermitian
$H_{\rm eff}$. The non-Hermitian  operator
$H_{\rm eff}$,  Eq.  (\ref{heff}), appears only at an intermediate stage 
of the FPO formalism. It is characteristic of the subsystem 
localized in a certain space region  and opened by coupling
it to the surrounding subspace of extended scattering states. 
It contains the time operator. The observables related to the whole system, 
such as the resonance structure of the scattering process,
are described by the wave functions 
$\Psi^E$ that are solutions of the equations (\ref{Psi}) and  (\ref{PsiF}),
respectively, with the Hermitian Hamilton operator $H$.
The unitarity of the $S$ matrix is guaranteed at all energies.

Although $H_{\rm eff}$ is an operator appearing only at an intermediate 
stage, it causes fundamental phenomena  involved in the FPO formalism.
The $\Psi^E$ can be represented in the set of eigenfunctions $\phi_\lambda$
of $H_{\rm eff}$ which are biorthogonal, see Sect. II.
This nontrivial representation causes, among others, 
the time asymmetry involved in the FPO formalism: 
the asymmetry rests on the fact that only localized states decay,
i.e. states for which the representation   (\ref{total1}) is meaningful.
This fact corresponds to the formulation  by Bohm et al. \cite{bohm}
that the states have to be prepared before they can be registered. 
Furthermore, the
appearance of the non-Hermitian Hamilton operator $H_{\rm eff}$ in the FPO
formalism guarantees the unified description of resonance and decay phenomena.
Its eigenvalues $z_\lambda$
describe, on the one hand, the resonance phenomena involved in
the resonance part of the $S$ matrix, Eq. (\ref{smatr}). On the other hand,
the decay of the states $\lambda$ lying at the energy
Re$\,(z_\lambda)$ is determined by Im$\,(z_\lambda)$. Thus, the resonance 
phenomena are directly related to the decay properties of the system.
This result being in accordance with longtime experience, 
as well as with the time asymmetry
can not be obtained in standard quantum mechanics with Hermitian Hamilton
operators in the Hilbert space. 

As a result of the study on the basis of the FPO formalism, we state that 
the basic relation (\ref{tauga}) between life time and decay width of resonance
states holds not only at low level density (where the resonances
are well separated from one another)
but also at high level density (where trapped long-lived resonance states are
superposed by a smooth background). In these cases the 
decay rate $k_{\rm \lambda}(t)$ of the individual state 
$\lambda$ and  the average decay rate $k_{\rm av}(t)$ of the long-lived 
trapped resonance states, respectively,  is  (almost) constant in time. 
Deviations may occur due to the position of thresholds in the neighborhood.
In the overlapping regime however, the mutual influence of the different 
resonance states onto each other causes a time dependence of 
the decay rate: $k_{\rm gr}(t)$ decreases monotonously with 
increasing $t$ according to 
(\ref{tdse4}). In the case the decay rate is constant in time, the
decay occurs according to an exponential law while the decay 
takes place according to a non-exponential law
when the resonances are not well separated from one another and the decay rate
depends on time.

\vspace{.3cm}

{\bf Acknowledgments:} Valuable discussions with A. Bohm are gratefully
acknowledged.

\end{document}